\documentstyle[emulateapj]{article}

\slugcomment{Accepted 1998 November for Astrophysical Journal Letters}
\lefthead{Drinkwater et al.}
\righthead{Unresolved Compact Galaxies}

\def\Bj{\hbox{$B_J$}}

\def\Mb{\hbox{{\rm M}$_B$}}
\def\Mpc{{\rm\thinspace Mpc}}
\def\asec{{\rm\thinspace arcsec}}
\def\deg{{\rm\thinspace deg}}
\def\degsq{\hbox{$\deg^{2}\,$}}
\def\pdegsq{\hbox{$\deg^{-2}\,$}}
\def\mag{{\rm\thinspace mag}}

\def\Bu{\hbox{$B\mag\asec^{-2}\,$}}
\def\s{{\rm\thinspace s}}
\def\km{{\rm\thinspace km}}
\def\kms{\hbox{$\km\s^{-1}\,$}}
\def\kmsmpc{\hbox{$\km\s^{-1}\Mpc^{-1}\,$}}
\def\mpc3{\hbox{$\Mpc^{-3}\,$}}
\def\kpc{{\rm\thinspace kpc}}

\def\one_wide{8.5cm}
\def\two_wide{14.0cm}
\def\nd{\nodata}
\def\caii{{\rm Ca\hbox{\small II}}}

\def\ha{{\rm H\hbox{$\alpha$}}}
\def\hb{{\rm H\hbox{$\beta$}}}
\def\hd{{\rm H\hbox{$\delta$}}}
\def\he{{\rm H\hbox{$\epsilon$}}}
\def\hii{H\,{\small II}}
\def\nii{{\rm N\hbox{\small II}}}

\def\oiii{{\rm O\hbox{\small III}}}

\begin{document}

\title{The Fornax Spectroscopic Survey:
The Number of Unresolved Compact Galaxies}

\author{Michael J. Drinkwater\altaffilmark{1,7},
Steven Phillipps\altaffilmark{2},
Michael D. Gregg\altaffilmark{3,7}, 
Quentin A. Parker\altaffilmark{4}, Rodney M. Smith\altaffilmark{5},\nl
Jonathan I. Davies\altaffilmark{5}, J. Bryn Jones\altaffilmark{2},
Elaine M. Sadler\altaffilmark{6}}

\altaffiltext{1}{School of Physics, University of New South Wales, Sydney 2052, Australia}
\altaffiltext{2}{Department of Physics, University of Bristol, Tyndall Avenue, Bristol BS8 1TL, UK}
\altaffiltext{3}{University of California, Davis, and IGPP, Lawrence Livermore National Laboratory,
L-413, Livermore, CA 94550, USA}
\altaffiltext{4}{Anglo-Australian Observatory, Coonabarabran, New South Wales,
2357, Australia}
\altaffiltext{5}{Department of Physics and Astronomy, University of Wales Cardiff, PO Box 913, Cardiff CF2 3YB, UK}
\altaffiltext{6}{School of Physics, University of Sydney, New South Wales 2006, Australia}
\altaffiltext{7}{Visiting Astronomer, Cerro Tololo
InterAmerican Observatory, National Optical Astronomy Observatory}

\begin{abstract}
We describe a sample of thirteen bright ($18.5<\Bj<20.1$) compact
galaxies at low redshift ($0.05<z<0.21$) behind the Fornax
Cluster. These galaxies are unresolved on UK Schmidt sky survey
plates, so would be missing from most galaxy catalogs compiled from
this material. The objects were found during initial observations of
{\it The Fornax Spectroscopic Survey}. This project is using the
Two-degree Field spectrograph on the Anglo-Australian Telescope to
obtain spectra for a complete sample of {\it all} 14000 objects, {\it
stellar and non-stellar}, with $16.5 < B_J < 19.7$, in a 12 square
degree area centered on the Fornax cluster of galaxies.
The surface density of compact galaxies with magnitudes
$16.5<\Bj<19.7$ is $7\pm3\pdegsq$, representing $2.8\pm1.6\%$ of all
local ($z<0.2$) galaxies to this limit. There are $12\pm3\pdegsq$ with
$16.5<\Bj<20.2$. They are luminous ($-21.5<\Mb<-18.0$, for
$H_0=50\kmsmpc$) and most have strong emission lines (\ha\ equivalent
widths of 40--200\,\AA) and small sizes typical of luminous \hii\
galaxies and compact narrow emission line galaxies. Four out of
thirteen have red colors and early-type spectra, so are unlikely to
have been detected in any previous surveys.
\end{abstract}


\section{Introduction}
\label{sec_intro}

Galaxy detection in many optical surveys, especially those
based on photographic data, suffers strong selection effects as a
function of surface brightness. The difficulty of detecting low
surface brightness galaxies is well accepted (\cite{imp88,fer95}), but
at the other extreme, it has been argued that there is {\em no} strong
selection against high surface brightness galaxies
(\cite{all79,kru87}).  Most galaxy surveys based on photographic
material ($\Bj<21$) have assumed---implicitly---that very few, if any,
galaxies are unresolved (e.g.\ \cite{mad90a}).  Morton, Krug \&
Tritton (1985) attempted to check this, taking spectra of all 606
stellar objects brighter than $B=20$ in an area of 0.31\degsq\ but
found no galaxies.  Colless et al.\ (1991) found seven galaxies among
a sample of 117 faint compact objects, but these were so faint
($\Bj\simeq22.5$) that the image classifications were not conclusive.

Many unresolved galaxies have been found in QSO surveys
(\cite{dow81,kk88,boy91}). More recently the Edinburgh-Cape blue
object survey (\cite{sto97}) and the Anglo-Australian Observatory 2dF
QSO redshift survey (\cite{boy98}) have produced further examples.
Many compact galaxies have also been found among \hii\ galaxies 
in objective prism surveys: some 50\% of these have starlike
morphology (\cite{mel87}). The compact narrow emission line galaxies
(CNELGs) found in the Koo \& Kron (1988) survey have been studied in
detail (\cite{koo94,koo95,guz96,guz98}): 35 have been found in an area
of 1.2\degsq\ to a magnitude limit of $\Bj=22.5$. These are
very blue with luminosities, scale sizes and emission line
spectra typical of nearby luminous \hii\ galaxies
(cf.\ \cite{sal89,ter91,gal97}). Similar galaxies have been found at
higher ($0.4<z<1$) redshifts (\cite{phi97}) and their distribution may
even extend to $z\approx3$ (\cite{low97}).

In this Letter we describe a new sample of bright ($\Bj\lesssim20.1$),
compact galaxies unresolved on the photographic sky survey plates
commonly used to create galaxy catalogs. Unlike previous work, this
sample is from a complete spectroscopic survey of {\em all} objects in
an area of sky, so we can estimate the fraction of all galaxies which
are compact. A population of compact galaxies missing in normal galaxy
surveys (e.g. \cite{col98}) would be important for several reasons
(\cite{sch94}).

\section{The Fornax Spectroscopic Survey}
\label{sec_obs}

{\em The Fornax Spectroscopic Survey} (see \cite{fss1} for details) is
designed to provide a census of galaxies in the local Universe free of
morphological selection criteria. We are using the Two-degree Field
spectrograph (2dF) on the Anglo-Australian Telescope to obtain spectra
for all 14,000 objects, stellar and non-stellar, in four 2dF fields
($12.5\degsq$) centered on the Fornax Cluster, with magnitude limits
of $16.5<\Bj<19.7$ (and somewhat deeper for unresolved images). 
Our targets are drawn from a UK Schmidt $B_{J}$ sky survey plate
centered on the Fornax Cluster (\cite{phi87}) digitized by the
Automated Plate Measuring facility (APM) (\cite{irw94}).  Although we
observe objects of all morphological types, we used the automated APM
classifications of the objects as ``stellar'' (probably stars) or
``resolved'' (probably galaxies) to optimize our photographic
photometry. The magnitudes of the resolved objects were measured by
fitting exponential intensity profiles to the run of area against
isophotal threshold in the APM data (\cite{dav88,dav90}).  The stellar
$B_{J}$ magnitudes were taken from the APM catalog data (\cite{irw94})
which uses internal self-calibration to fit stellar profiles,
correcting for the non-linear response of the photographic emulsion.

Here
we present preliminary results from the first field centered at
$\alpha = 03^{h}38^{m}29^{s}, \delta = -35\arcdeg 27\arcmin 01\arcsec$
(J2000) observed in 
Semesters 1996B and 1997B. We
observed and identified 992 (77\%) of the resolved objects to a limit
of $\Bj<19.7$, 675 (38\%) of stellar objects to the same limit and a
total of 1112 (43\%) of the stellar objects to the deeper limit of
$\Bj<20.2$. Figure~\ref{fig_histo} shows the completeness of our
observations as a function of magnitude and color. Our main result is
that thirteen of the ``stellar'' objects have recession velocities of
14,000--60,000\kms\ (see
Table~\ref{tab_members}). These galaxies are well beyond the Fornax
Cluster ($v\simeq 1500\kms$) and most (nine) have strong emission line
spectra.

\placefigure{fig_histo}

\placetable{tab_members}

\section{Properties of the new galaxies}
\label{sec_props}

In Figure~\ref{fig_mags} we compare the distribution of the new compact
galaxies to previously detected CNELGs (\cite{koo94,koo95}) in
magnitude-redshift space.  There is considerable overlap in absolute
magnitude, but as expected from a larger area survey with a
brighter magnitude limit, our galaxies occupy a region in this diagram
at lower redshift and brighter apparent magnitude.

\placefigure{fig_mags}

The compact nature of the new galaxies prohibits a detailed analysis
of their scale sizes and central surface brightness using our imaging
data. The galaxy images are unresolved on photographic sky survey
plates ($1\farcs5$ seeing FWHM). We therefore estimate conservative
upper limits (not correcting for photographic saturation which occurs
at about 21\Bu) to their scale lengths to be $\simeq 1''$ (assuming an
image FWHM of $1\farcs5$). This upper limit has been confirmed by a
CCD image of one of the galaxies taken with the CTIO\footnote{CTIO is
operated by the Association of Universities for Research in Astronomy
Inc. (AURA), under a cooperative agreement with the National Science
Foundation as part of the National Optical Astronomy Observatories.}
1.5m Telescope which was only marginally resolved in $1\farcs2$
seeing. At the range of distances indicated, this corresponds to
physical scale sizes of 1--4\kpc\footnote{We adopt $H_0=50\kmsmpc$ and
$q_0=0.1$.}, somewhat smaller than local spiral galaxies
(\cite{jon96}) and at least as small as CNELGs and luminous \hii\
galaxies (\cite{phi97}).  Despite the small scales, these galaxies are
not dwarfs in terms of their luminosities which are within a factor 10
or so of $L_{*}$; indeed some of them exceed $L_{*}$. This is because
of their high surface brightnesses: the scale size limits of $1''$
imply central surface brightnesses 19--21\Bu, as bright as the CNELGs
and the luminous \hii\ galaxies (\cite{phi97}).

We obtained Cousins BVI CCD images of our survey region using the CTIO
Curtis Schmidt Telescope. The low spatial resolution (3\arcsec\ FWHM)
leaves the compact galaxies unresolved but the data allow us to
calculate photometry and aperture ($8''$ radius) colors.  The
K-corrected $B-V$ colors (Table \ref{tab_members}) of the
emission-line compact galaxies place them among the CNELG and \hii\
galaxies, and are consistent with relatively high recent star
formation rates (\cite{lar78}).

The nine emission-line compact galaxies all have strong narrow \ha\
lines: none are resolved at our resolution of 9\,\AA\ or 450\kms. The
\ha\ rest equivalent widths listed in Table~\ref{tab_members} nearly
all exceed the mean value for our overall background sample of
emission line objects of $EW(\ha)\simeq37$\,\AA, typical of local
spirals (\cite{ken92}). The values for the compact galaxies,
$\sim$30--190\,\AA\ are more like those seen in low redshift \hii\
galaxies (\cite{gal97}). Given the range of overall sizes of these
objects, it is interesting to consider what Cowie et al.\ (1996) call
the stellar mass doubling time, i.e.\ the time it would take for the
current star formation rate (SFR) to double the existing underlying
stellar mass. This can be derived directly from the equivalent widths:
Cowie et al.\ note that $EW(\ha) \simeq 60$\,\AA\ separates
galaxies undergoing rapid star formation, with mass doubling times
less than $10^{10}$ years, from those with moderate SFRs which can be
maintained for a Hubble time. At the highest SFR, Cowie et al.\ find
that a mass doubling time of $2 \times 10^{9}$ years empirically
corresponds to their galaxies with $EW(\ha) \simeq 125$\,\AA. Our
fastest star formers, like J0338$-$3545, should have mass doubling
times of this order.

In Figure~\ref{fig_ratio} we show the [\oiii]/\hb\ vs.\ [\nii]/\ha\
emission line ratio diagram for our new galaxies compared to the
Gallego et al.\ (1997) emission line sample and the Koo et al.\
CNELGs. This shows that new compact galaxies are actively star-forming
as they closely follow the general \hii\ region relationship. They
display generally high excitation as measured by [\oiii]/\hb, putting
them in the H{\small II}H class more than the starburst nucleus class as
defined by Gallego et al. The new compact galaxies have very similar
properties to the CNELGs. We do not draw any conclusion from the lack
of low excitation objects: this may be a selection effect as we are
only considering the unresolved galaxies in our survey in this
sample. Similar conclusions can be drawn from a plot of the
excitation against absolute magnitude.

\placefigure{fig_ratio}

We also found four compact galaxies which did not have strong emission
lines and are therefore not shown in the excitation diagrams. These
are unlikely to have been detected in previous work on compact
galaxies because of their weak line emission and generally redder
colors which would exclude them from most QSO surveys.  One of them,
J0339$-$3547, has a post starburst spectrum with strong Balmer
absorption lines. We have used the two ratios of absorption feature
strengths \caii H+\he/\caii K and \hd/Fe{\small I}4045 to estimate the
age of the galaxy since the end of the starburst (\cite{leo96}). For
J0339$-$3547 these ratios are 0.89 and 0.69 respectively. For the
Leonardi \& Rose model of a starburst lasting 0.3\,Gy, this is
indicative of a very strong starburst about 1\,Gy after the end of the
burst. This galaxy may represent an intermediate stage between the
CNELG types and the dwarf spheroidal remnants proposed by Koo et al.\
(1995).  By comparison, J0340$-$3510 has ratios of 1.12 and 0.97, and
a composite spectrum of 60 normal early type galaxies from the survey
has values of 1.10 and 0.95, both consistent with the Leonardi \& Rose
values for an old population.

\section{Numbers of compact galaxies}
\label{sec_disc}

To estimate the true numbers of compact galaxies from our sample we
must first make a completeness correction. The color distributions in
Figure~\ref{fig_histo} show that the compact galaxies are all bluer
than $B_J-R_F=1.6$, so the best correction can be taken from the
fraction of blue ($B_J-R_F<1.6$) stellar objects observed; to
$B_J=19.7$ we observed 31\% of the blue stellar objects, so the
corrected number of compact galaxies to this limit is
$7\div0.31=23\pm9$, equivalent to a surface density of
$7\pm3\pdegsq$. The surface density to $B_J=20.2$ (completeness of
35\%) is $12\pm3\pdegsq$.

We can use our observations of field galaxies to estimate the fraction
of normal galaxies represented by the compact galaxies.  At redshifts
$z>0.2$, the \ha\ line is shifted out of our 2dF spectra and our
galaxy sample is less complete, so we define a ``local'' comparison
field sample to be all galaxies in the field beyond the Fornax Cluster
but at redshifts $z<0.2$. We successfully observed 992 resolved
objects to $B_J = 19.7$, of which 675 were ``local'' background
galaxies: this is the minimum number of local field galaxies.  There
are $1296-992=304$ resolved objects still to observe, so the maximum
number of local field galaxies is $675+304=979$.  The number of local
($z<0.2$) background galaxies to our limit is
$827\pm152$. The expected 23 compact galaxies among the stellar
objects therefore constitute $2.8\pm1.6\%$ of the local galaxy
population. These would be missed by any surveys of objects classified
by the APM as ``galaxies'' from UK Schmidt photographic data with
$16.5<\Bj<19.7$. These selection criteria are typical of previous
surveys (\cite{mad90b,col98}).

This conclusion is for a magnitude, not volume, limited sample, but in
fact the compact galaxies do occupy a similar volume of space to the
general run of galaxies to this magnitude limit. For instance, the 2dF
galaxy redshift survey, limited at a very similar $B_{J}$ to ours, has
a mean redshift of 33000 km s$^{-1}$ (\cite{col98}), close to the mean
of the compact galaxies.  The galaxy catalog used for that survey has
a mean surface density of 222\pdegsq\ at $\Bj<19.7$ (M. Colless,
private communication), for which the compact galaxies would represent
an additional $3.2\pm1.2\%$.

Only about half of the local galaxy sample exhibits significant
emission line features, so the new compact galaxies constitute a
larger fraction of emission line galaxies (3--5\%). They contribute an
even larger fraction of strong \ha\ emitters with $EW(\ha) >
40$\,\AA, so may make a small but measurable contribution to the local
star formation rate.

The new compact galaxies have very similar absolute magnitudes, sizes
and (in most cases) emission line properties to the Koo et al.\
(1994,1995) CNELGs. The distributions shown in Figure~\ref{fig_mags}
suggest that they are a continuation of the CNELGs to lower
redshifts and brighter apparent magnitudes.  A better way to compare
these populations is by the volume density. Koo et al.\ (1994) derive
a CNELG density of $7.5\times10^{-5}$\mpc3\ compared to the value of
$17\times10^{-5}$\mpc3\ for local H{\small II}H plus DH{\small II}H
galaxies (Salzer et al.\ 1989). For our sample of compact galaxies,
using the $1/V_{max}$ method we obtain a similar value of
$(13\pm4)\times10^{-5}$\mpc3. These results are consistent with the
respective galaxy populations being related, but we prefer not to draw
any conclusions until we can analyze the compact galaxies in the
context of our complete sample.

\acknowledgments

We thank our referee for detailed suggestions which greatly improved
the presentation of this work. We also thank Dr.\ Lewis Jones for
helpful discussions and Dr.\ Jesus Gallego for providing data for
Figure~\ref{fig_ratio}. SP acknowledges the support of the Royal
Society via a University Research Fellowship. JBJ is supported by the
UK PPARC. Part of this work was done at the Institute of Geophysics
and Planetary Physics, under the auspices of the U.S. Department of
Energy by Lawrence Livermore National Laboratory under contract
No.~W-7405-Eng-48.



\figcaption{Histograms showing the completeness of our observations as
functions of magnitude and color for stellar and resolved objects.
The colors were taken from the APM catalog using magnitudes derived
from stellar profile fitting, so are only indicative for the
resolved objects. In each case the upper histogram is the total number of
objects and the lower gives the number observed and identified. The
triangles indicate the locations of the new compact galaxies.
\label{fig_histo}}

\figcaption{Distribution of absolute and apparent magnitudes of the
new compact galaxies (triangles) compared to the Koo et al.\
(1994,1995) CNELGs (crosses) as a function of redshift.
\label{fig_mags}}

\figcaption{Emission-line diagnostic diagram of [\oiii]/\hb\ vs.\
[\nii]/\ha. The new compact galaxies (triangles) are compared
with CNELGs (crosses) and a range of local galaxies from the UCM
survey (\cite{gal97}).
\label{fig_ratio}}


\begin{deluxetable}{crrrrrrrrrr}
\scriptsize
\tablecaption{Properties of the Compact Galaxies\label{tab_members}}
\tablewidth{0pt}
\tablehead{
\colhead{RA (J2000) Dec}&\colhead{$z$}&
\colhead{$V$} & \colhead{\bv} & \colhead{$V-I$} &
\colhead{\Bj}& \colhead{$\bv_0$\tablenotemark{a}} &\colhead{\Mb\tablenotemark{a}} &
\colhead{\underline{$[OIII]$}}& 
\colhead{\underline{$[NII]$}}& 
\colhead{W$_{\ha}$}
\\
\colhead{} & \colhead{}  &
\colhead{\footnotesize mag} & \colhead{\footnotesize mag} & \colhead{\footnotesize mag}&
\colhead{\footnotesize mag} &\colhead{\footnotesize mag} &\colhead{\footnotesize mag} &
\colhead{\hb} &\colhead{$\ha$} & \colhead{\footnotesize \AA}
}
\startdata
3:34:45.47 $-$35:38:18.0& 0.0453&18.73$\pm$0.08&0.52$\pm$0.10&0.48$\pm$0.14&19.1&0.43&$-$18.0&4.0 & 0.12 & 39 \nl
3:34:53.03 $-$36:03:03.5& 0.2130&19.20$\pm$0.08&1.00$\pm$0.16&0.83$\pm$0.15&19.9&0.57&$-$21.1&2.5 & 0.21 & 40 \nl
3:35:33.06 $-$35:01:12.8& 0.1593&18.12$\pm$0.05&0.55$\pm$0.06&0.90$\pm$0.07&18.5&0.55&$-$21.5&1.8 & 0.25 &133 \nl
3:38:56.50 $-$35:45:00.3& 0.1157&19.42$\pm$0.09&0.22$\pm$0.12&0.71$\pm$0.16&19.6&0.22&$-$19.8&4.6 & 0.12 &189 \nl
3:39:18.37 $-$35:32:40.7& 0.1838&18.40$\pm$0.04&0.93$\pm$0.10&1.27$\pm$0.05&19.1&0.55&$-$21.5&\nd & 0.42 &  6 \nl
3:39:51.33 $-$35:47:52.8& 0.1553&19.09$\pm$0.06&0.86$\pm$0.14&1.23$\pm$0.08&19.7&0.55&$-$20.4&\nd &\nd   &\nd \nl
3:40:06.66 $-$36:04:27.1& 0.1156&18.79$\pm$0.05&0.74$\pm$0.09&0.80$\pm$0.09&19.3&0.51&$-$20.6&\nd & 0.39 & 59 \nl
3:40:25.54 $-$34:58:33.5& 0.1039&19.11$\pm$0.07&0.47$\pm$0.12&0.90$\pm$0.15&19.4&0.47&$-$19.6&\nd &\nd   &  8 \nl
3:40:57.25 $-$35:10:34.1& 0.1616&19.08$\pm$0.06&1.30$\pm$0.20&1.00$\pm$0.10&20.0&0.84&$-$20.4&\nd &\nd   &\nd \nl
3:41:32.89 $-$35:20:08.4& 0.0778&19.73$\pm$0.18&0.50$\pm$0.20&0.65$\pm$0.28&20.1&0.34&$-$18.3&3.29& 0.13 & 44 \nl
3:41:56.94 $-$35:44:01.0& 0.1163&19.71$\pm$0.12&0.61$\pm$0.21&0.82$\pm$0.20&20.1&0.39&$-$19.2&4.01& 0.11 & 31 \nl
3:41:59.59 $-$35:09:01.2& 0.1391&18.84$\pm$0.06&0.80$\pm$0.11&1.05$\pm$0.09&19.4&0.53&$-$21.0&1.72& 0.34 &136 \nl
3:42:38.68 $-$35:56:21.9& 0.1070&19.42$\pm$0.09&0.58$\pm$0.14&0.92$\pm$0.12&19.8&0.37&$-$19.3&3.10& 0.21 &138 \nl
\enddata
\tablenotetext{a}{K-corrected (\cite{col80}) using $H_0=50\kmsmpc$ and $q_0=0.1$.}
\end{deluxetable}

\clearpage


\epsscale{0.5}
\plotone{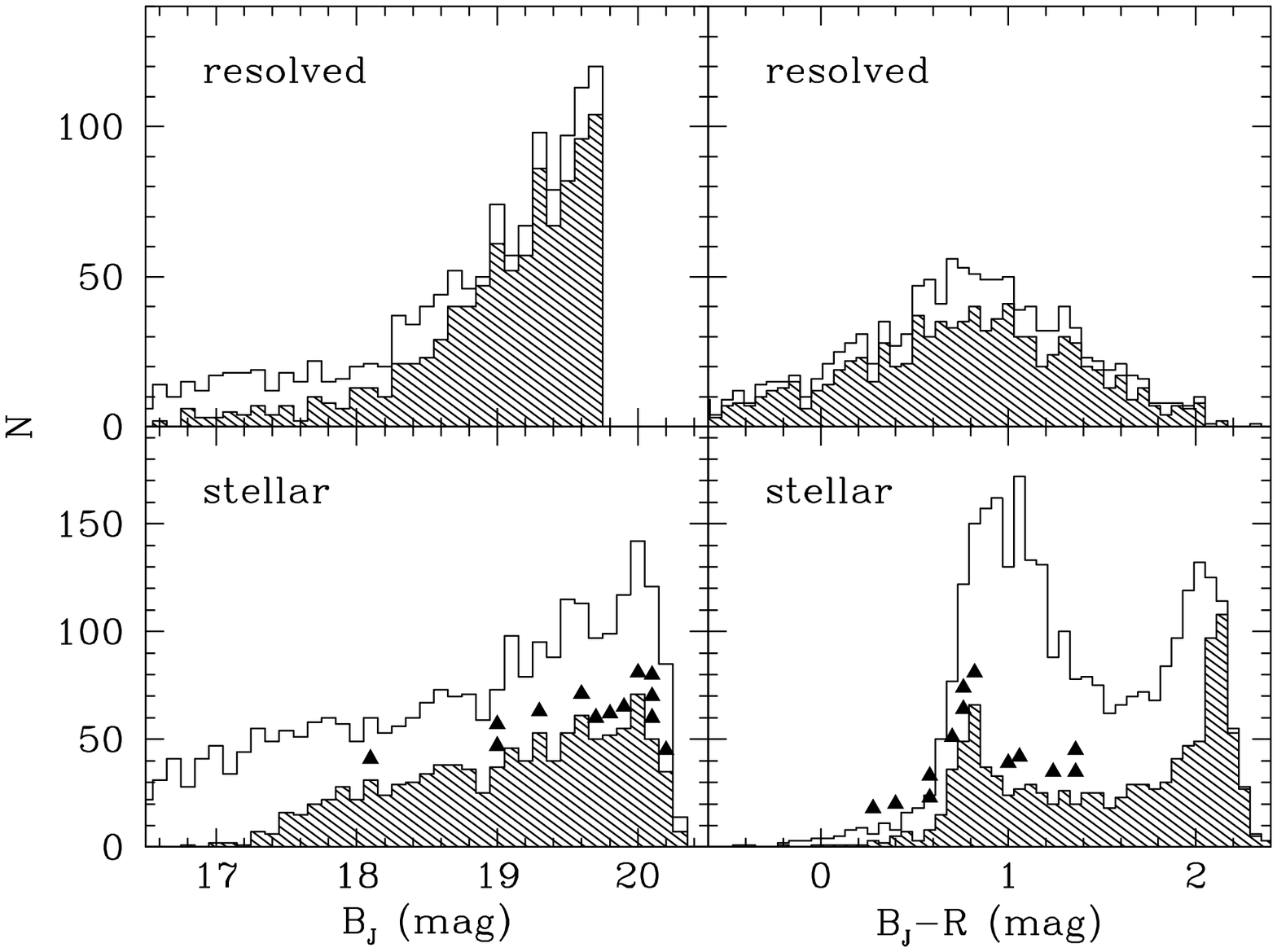}

\epsscale{0.5}
\plotone{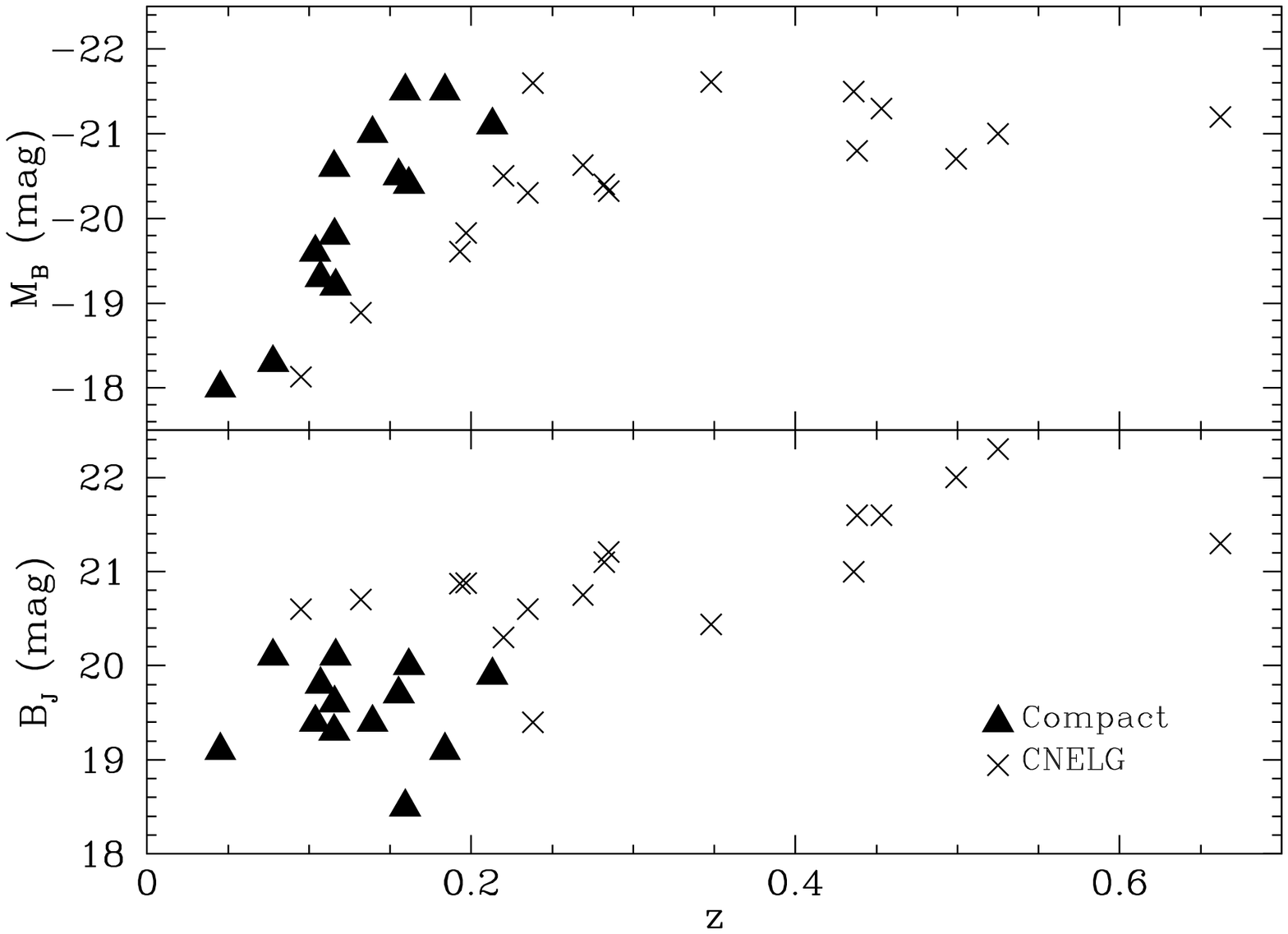}

\epsscale{0.5}
\plotone{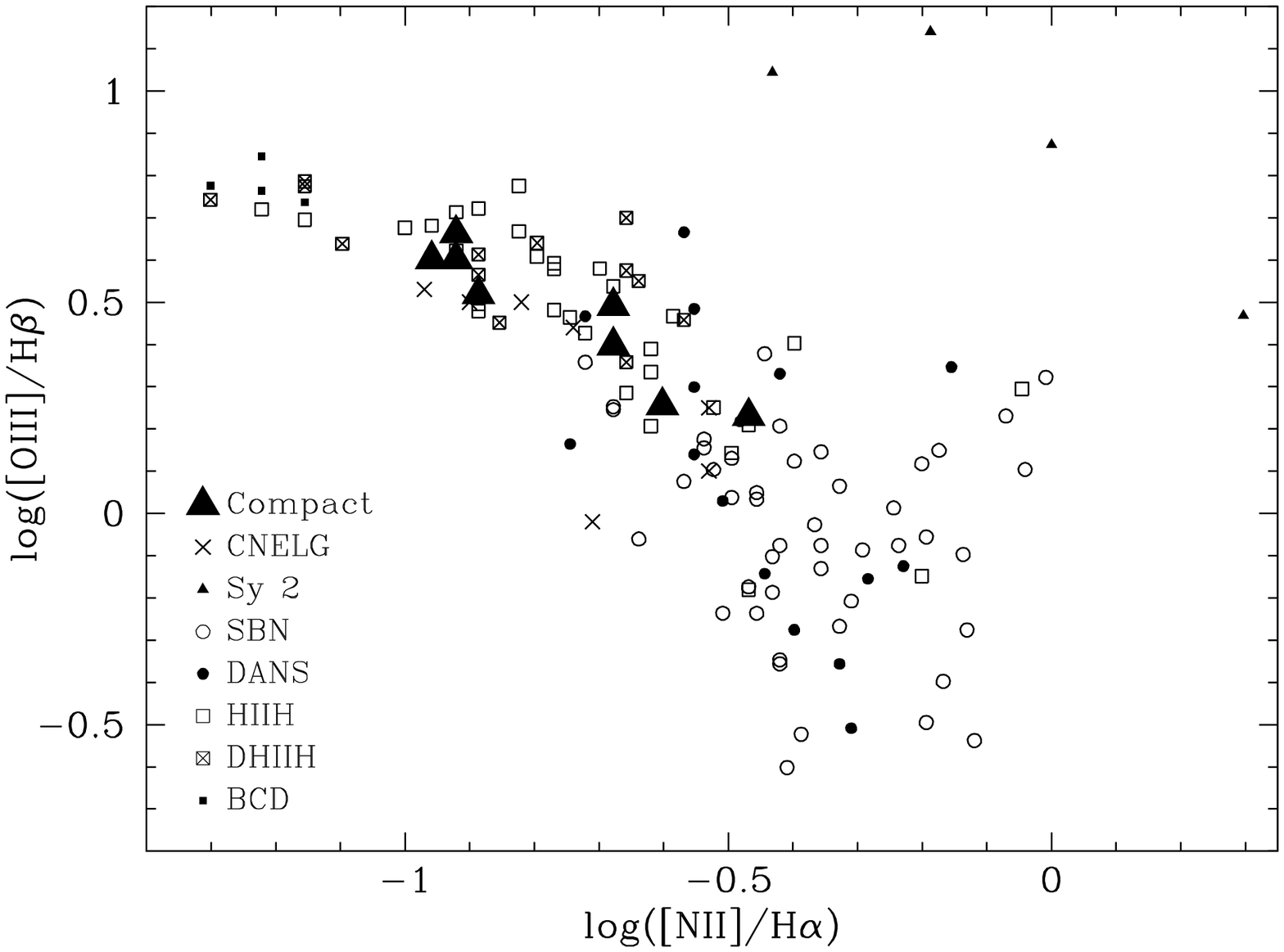}


\begin{thebibliography}{}
\bibitem[Allen \& Shu 1979]{all79} Allen, R. J., Shu, F. H. 1979, \apj, 227, 67
\bibitem[Boyle et al.\ 1991]{boy91} Boyle, B.J., Jones, L.R., Shanks, T. 1991, \mnras, 251, 482
\bibitem[Boyle et al.\ 1998]{boy98} Boyle, B.J., Smith, R.J., Shanks, T., Croom, S.M., Miller, L., Read, M. 1998, in IAU Symp. 183, Cosmological Parameters and Evolution of the Universe, in press
\bibitem[Coleman et al.\ 1980]{col80} Coleman, G.D., Wu, C.-C., Weedman, D.W. 1980, \apjs, 43, 393
\bibitem[Colless 1998]{col98} Colless, M. 1998, Phil.Trans.R.Soc.Lond.A, in press
\bibitem[Colless et al.\ 1991]{col91} Colless, M., Ellis, R.S., Taylor, K., Shaw, G. 1991, MNRAS, 253, 686
\bibitem[Cowie et al.\ 1996]{cow96} Cowie, L.L., Songaila, A., Hu, E.M., Cohen, J.G. 1996, AJ, 112, 839
\bibitem[Davies 1990]{dav90} Davies, J.I. 1990, MNRAS, 245, 350
\bibitem[Davies et al.\ 1988]{dav88} Davies, J.I., Phillipps, S., Cawson, M.G.M., Disney, M.J., Kibblewhite, E.J. 1988, MNRAS, 232, 239
\bibitem[Downes \& Margon, 1981]{dow81} Downes, R.A., Margon, B. 1981, \aj, 86, 19
\bibitem[Drinkwater et al.\ 1998]{fss1} Drinkwater, M.J., Phillipps, S., Davies, J.I., Gregg, M.D., Jones, J.B., Parker, Q.A., Sadler, E.M., Smith, R.M. 1998, \mnras, in preparation
\bibitem[Ferguson \& McGaugh 1995]{fer95} Ferguson, H.C., McGaugh, S.S. 1995, AJ, 440, 470
\bibitem[Gallego et al.\ 1997]{gal97} Gallego, J., Zamorano, J., Rego, M., Vitores, A.G. 1997 \apj, 475, 502
\bibitem[Guzm\'an et al.\ 1996]{guz96} Guzm\'an, R., Koo, D.C., Faber, S.M., Illingworth, G.D., Takamiya, M., Kron, R.G., Bershady, M.A. 1996, ApJ, 460, L9
\bibitem[Guzm\'an et al.\ 1998]{guz98} Guzm\'an, R., Jangren, A., Koo, D.C., Bershady, M.A., Simard, L. 1998, ApJ, 495, L13
\bibitem[Impey, Bothun \& Malin 1988]{imp88} Impey, C., Bothun, G., Malin, D. 1988, ApJ, 330, 634
\bibitem[Irwin, Maddox \& McMahon, 1994]{irw94} Irwin, M., Maddox, S., McMahon, R. 1994, Spectrum, 2, 14
\bibitem[de Jong 1996]{jon96} de Jong, R.S. 1996, A\&A, 313, 45
\bibitem[Kennicutt 1992]{ken92} Kennicutt, R.C. 1992, ApJ Suppl., 79, 255
\bibitem[Koo \& Kron 1988]{kk88} Koo, D.C., Kron, R.G. 1988, \apj, 325, 92
\bibitem[Koo et al.\ 1994]{koo94} Koo, D.C., Bershady, M.A., Wirth, G.D., Stanford, S.A., Majewski, S.R. 1994, ApJ, 427, L9
\bibitem[Koo et al.\ 1995]{koo95} Koo, D.C., Guzm\'an, R., Faber, S.M., Illingworth, G.D., Bershady, M.A., Kron, R.G., Takamiya, M. 1995, ApJ, 440, L49
\bibitem[van der Kruit 1987]{kru87} van der Kruit, P.C. 1987, A\&A, 173, 59
\bibitem[Larson \& Tinsley 1978]{lar78} Larson, R.B., Tinsley, B.M. 1978, ApJ, 219, 46
\bibitem[Leonardi \& Rose 1996]{leo96} Leonardi, A.J., Rose, J.A. 1996, \aj, 111, 182
\bibitem[Lowenthal et al.\ 1997]{low97} Lowenthal, J.D., Koo, D.C., Guzm\'an, R., Gallego, J., Phillipps, A.C., Faber, S.M., Vogt, N.P., Illingworth, G.D., Gronwall, C. 1997, \apj, 489, 543
\bibitem[Maddox et al.\ 1990a]{mad90a} Maddox, S.J., Sutherland, W.J., Efstathiou, G., Loveday, J. 1990a, MNRAS, 243, 692
\bibitem[Maddox et al.\ 1990b]{mad90b} Maddox, S.J., Efstathiou, G., Sutherland, W.J. 1990b, MNRAS, 246, 433
\bibitem[Melnick 1987]{mel87} Melnick, J., 1987, in Starbursts and Galaxy Evolution, ed. T.X. Thuan, T. Montmerle, \& J. Tran Thang Van (Paris: Editions Fronti\`eres), 215
\bibitem[Morton, Krug \& Tritton 1985]{mor85} Morton, D.C., Krug, P.A., Tritton, K.P. 1985, MNRAS, 212, 325
\bibitem[Phillipps et al.\ 1987]{phi87} Phillipps, S., Disney, M.J., Kibblewhite, E.J., Cawson, M.G.M. 1987, MNRAS, 229, 505
\bibitem[Phillips et al.\ 1997]{phi97} Phillips, A.C., Guzm\'an, R., Gallego, J., Koo, D.C., Lowenthal, J.D., Vogt, N.P., Faber, S.M., Illingworth, G.D., 1997, \apj, 489, 543
\bibitem[Salzer et al.\ 1989]{sal89} Salzer, J.J., MacAlpine, G.M., Boroson, T.A. 1989, ApJS, 70, 479
\bibitem[Schade \& Ferguson 1994]{sch94} Schade, D., Ferguson, H.C. 1994, MNRAS, 267, 889
\bibitem[Stobie et al.\ 1997]{sto97} Stobie, R., et al.\ 1997, MNRAS, 287, 848
\bibitem[Terlevich et al.\ 1991]{ter91} Terlevich, R., Melnick, J., Masegosa, J., Moles, M., Copetti, M.V.F. 1991, A\&AS, 91, 285
\end{thebibliography}
\end{document}